\documentclass{llncs}
\usepackage{epsf}

\begin{document}
\title{Components and Interfaces of a Process Management System
  for Parallel Programs\thanks{This
work was supported by the Mathematical, Information, and
Computational Sciences Division subprogram of the Office of
Advanced Scientific Computing Research, U.S.\ Department of Energy,
under Contract W-31-109-Eng-38.}}
\author{Ralph Butler\inst{1}  \and
  William Gropp\inst{2} \and
  Ewing Lusk\inst{2} }
\institute{Middle Tennessee State University \and Argonne National Laboratory}
\maketitle

\begin{abstract}
  Parallel jobs are different from sequential jobs and require a different
  type of process management.  We present here a process management system for
  parallel programs such as those written using MPI.  A primary goal of the
  system, which we call MPD (for multipurpose daemon), is to be scalable.  By
  this we mean that startup of interactive parallel jobs comprising thousands
  of processes is quick, that signals can be quickly delivered to processes,
  and that {\tt stdin}, {\tt stdout}, and {\tt stderr} are managed
  intuitively.  Our primary target is parallel machines made up of clusters of
  SMPs, but the system is also useful in more tightly integrated environments.
  We describe how MPD enables much faster startup and better runtime
  management of parallel jobs.  We show how close control of {\tt stdio} can
  support the easy implementation of a number of convenient system utilities,
  even a parallel debugger.  We describe a simple but general interface that
  can be used to separate any process manager from a parallel library, which
  we use to keep MPD separate from MPICH.
\end{abstract}

\section{Introduction}
\label{sec:intro}

A parallel programming environment may be viewed as comprising three
interacting components: a {\em job scheduler}, which decides what resources a
parallel job consisting of multiple processes will run on; a {\em process
  manager}, which starts and terminates processes and provides them
with a number of services; and a {\em parallel library} such as MPI, which a
parallel application calls upon for communications.  Since these components
need to communicate with one another, they are often integrated into a single
system.  An important research question is to what extent they
can be separated from one another with well-defined interfaces so that they
can be independently developed.  A further research question is whether the
resulting system can be made scalable to jobs involving thousands of
communicating processes.
In this paper we focus on the process
manager component.  We describe a design and an implementation we call MPD
(for multipurpose daemon) that provides both fast startup of parallel jobs
and a flexible run-time environment that supports parallel libraries through a
small, general interface.

A parallel job is both similar to a sequential job and different from one in
significant ways.  Resource allocation and scheduling are considerably more
complex, and we do not address those issues here.  But for process
management issues alone, complexity arises from the fact that there may be multiple
executables, multiple sets of command-line arguments, even different
environments for different processes.  Task farm jobs are different from true
parallel jobs in which processes will communicate with one another, not just
with a master process.  On clusters we must either set up all connections
ahead of time or provide some way for a process needing to establish
communication with another process to ask for help from the process manager
(who is the only one who knows where that other process is). 
The first approach is not scalable to large numbers of processes,
and in scalable applications not all such connections will eventually be
needed; therefore, a process manager must be able to provide information
services to parallel jobs to allow connections to be made dynamically.
Scalable startup is needed to make interactive parallel jobs feasible.
Parallel jobs also need scalable signal delivery and a reasonable semantics
for stdio redirection.

We assume familiarity with process management for sequential jobs on Unix.
Components of process management we take to be the process id, executable
name, environment variables, command-line arguments, signals (especially {\tt
  cntl-C}, {\tt cntl-Z}, and resume signals), {\tt stdin}, {\tt stdout}, and
{\tt stderr} and their redirection.  We differentiate process management from
{\em scheduling}, which is the problem of locating resources and a time to use
them.  Batch schedulers often combine scheduling with process management.

In Section~\ref{sec:related} we summarize related work.  In
Section~\ref{sec:design} we state our explicit design goals, how
these goals lead to implementation decisions, and interesting
features of the resulting system, including how it can be used to create a
parallel debugger out of an existing single-process debugger.
Section~\ref{sec:bnr} briefly describes a general-purpose
interface between process managers and parallel libraries and how this
interface is implemented on the process manager side by MPD.
Section~\ref{sec:experiments} summarizes preliminary experiments
that make us optimistic about the usefulness of MPD as a process manager for
large-scale systems.  We conclude with a summary of progress to date and a
description of our future plans.

The MPD system is in use and is available as open source as part of the MPICH
system~\cite{gropp-lusk-doss-skjellum:mpich}, obtainable from
{\tt http://www.mcs.anl.gov/mpi/mpich}.

An abbreviated report on this work appeared in~\cite{bgl00:mpd:pvmmpi00}.
Here we provide more detail than was possible there, describe new additions to
the system, and outline an interface that can be used by parallel programs to
obtain services from a process manager like MPD.

\section{Related Work}
\label{sec:related}

All parallel computing environments that support execution of truly parallel
programs (those in which any two processes can communicate
with one another) have had to address at least some of the issues that we
address with MPD.
Parallel programming systems, such as PVM~\cite{pvmbook}, p4~\cite{p4-paper},
and implementations of MPI such as MPICH~\cite{gropp-lusk-doss-skjellum:mpich}
and LAM~\cite{lam} all provide some mechanism for starting and running
parallel programs, often with a specialized daemon process.  MPD differs from
these systems in two ways.  First, it is independent of any particular
programming library, instead implementing a simple interface (described in
Section~\ref{sec:bnr}) by which any library, including these, can make use of
its services.  Second, it is designed specifically to enable rapid startup
of jobs consisting of hundreds to thousands of processes. 

Many systems are intended to manage a collection of computing resources
for both single-process and parallel jobs; see the survey by Baker et al.
\cite{baker96:clusters}.  
Typically, these use a daemon that manages
individual processes, with emphasis on jobs involving only a single process.
Widely used systems 
include PBS~\cite{pbs-home-page}, LSF~\cite{platform-home-page}, DQS
\cite{dqs-home-page}, and Loadleveler/POE~\cite{loadleveler}. 
The Condor
system~\cite{Litzkow88} is also widely used and supports parallel programs
that use PVM~\cite{Pruyne:1996:ICP} or MPI~\cite{Gropp:1998:MPI2Book,Snir:1998:MPI2Book}.  
More specialized systems, such as MOSIX~\cite{Barak:1993:MDO} and
GLUnix~\cite{Ghormley:1998:GGL}, provide single-system image support
for clusters.  
Harness~\cite{Beck99:harness,Migliardi:1999:PEH} shares with
MPD the goal of supporting management of parallel jobs.  Its primary research
goal is to demonstrate the flexibility of the ``plug-in'' approach to
application design, potentially providing a wide range of services.  The MPD system
focuses more specifically on the design and implementation of services
required for process management of parallel jobs, including high-speed startup
of large parallel jobs on clusters and scalable standard I/O management.
The book~\cite{foster99grid} provides a good overview of
metacomputing systems and issues, and Feitelson~\cite{feitelson95:survey} surveys support for scheduling parallel
processes.

\section{Design of MPD}
\label{sec:design}

In this section we describe our goals in constructing MPD and
outline the system's architecture.  

\subsection{Goals}
\label{sec:goals}

Several explicit goals have governed the design of the MPD system.
\begin{description}
\item[Simplicity]
The persistent (across jobs) part of the system should be simple and robust.
This part of the system should be runnable as root.  If its behavior
isn't completely transparent, we will never be able to convince system
administrators to run the daemons as root.
\item[Speed] Startup of parallel jobs should be quick enough to provide an
  interactive ``feel,'' so that large but short jobs make sense.  Large (in
  number of processes) but short (in time) characterizes system utilities such
  as those described in~\cite{SUT}.  Our immediate target is to start 1000
  processes in a few seconds, while still providing a way for such processes to
  establish contact with one another.  Our long-term goal is to support
  management of 10,000 processes.
\item[Robustness]
The persistent part of the system should be at least moderately
fault tolerant.  An unexpected crash of one machine should not bring down the
whole system.  There should be no single ``master'' process.
\item[Scalability]
The complexity or size of any component should not depend on the number of
components.
\item[Individual Process Environments]
It should be possible to start a parallel job in which the executable files,
environment variables, and command-line arguments are different for each
process.  It should be possible to collect return codes individually from
processes.
\item[Collective Identity of a Parallel Job] It should be possible to treat a
  parallel job as a single entity that can be suspended, continued (signaled,
  in general), or killed collectively as if it were a single process.  The
  system should manage {\tt stdin}, {\tt stdout}, and {\tt stderr} in a useful
  and scalable way and allow them to be redirected as if the parallel job were
  a single process.  An important component of a job's collective identity is
  its {\em termination}.  All resources allocated for the job, such as files,
  System V IPC's, other processes, etc., must be reliably freed, even if the
  job terminates abnormally.
\end{description}
It is explicitly not a goal of the MPD system to provide scheduling
services, which we believe to be a separate function from process management.
We expect the decision on precisely which resources to use to run a job to be
made by a separate scheduler, which will then communicate its decision to the
process management system.  Design of the interface by which this occurs is an
interesting problem, but not addressed here.  Note that many existing systems
combine scheduling and process management, an organization that we find limits
flexibility.  In this paper we focus solely on process management.

\subsection{Deriving the Design from the Goals}
\label{sec:deriving}

The goals of simplicity and robustness lead us to adopt a multicomponent
system.  The {\em daemon\/} itself is persistent (may run for weeks or months
at a time, starting many jobs), typically one instance per host in a
TCP-connected network.  {\em Manager\/} processes will be started by the
daemons to control the application processes ({\em clients\/}) of a single
parallel job and will provide most of the MPD features.  The goal of speed
requires that the daemons be in contact with one another prior to job startup,
and the goals of scalability and ``no master'' suggest that the daemons be
connected in a ring.\footnote{While a ring is not ultimately scalable, it is
  more so than the typical star used in many process management systems, and
  our experiments have shown it feasible for the thousand-daemon domain.} The
services that the managers will provide (see Section~\ref{sec:features})
suggest that they be in contact as well, and the fastest way for them to form
these connections is to inherit part of the ring connectivity of the daemons.
Separate managers for each user process support the individual process
environments.  The goal of having a collective identity for a parallel job
leads us to treat the {\tt mpirun} process as such a
representative and use it to deliver signals and {\tt stdin} to application
processes and collect {\tt stdout} and {\tt stderr} output from them.  This
suggests that the {\tt mpirun} process connect first to the daemon ring in
order to start the job, and then switch the connection to the manager ring in
order to control the job.  The goal of speed suggests that these latter
connections be restricted to a process running on the same host, either the
daemon itself or a persistent gateway process if the daemon is run as root, so
that authentication can be through the file system (a Unix rather than a
network socket).  We refer to all such processes as {\em console commands}.
The console commands {\tt mpd}, {\tt mpdtrace}, and {\tt mpdallexit} manage
the daemons themselves; {\tt mpdmpexec} and {\tt mpirun} start
parallel jobs; and {\tt mpdlistjobs}, {\tt mpdkilljob}, and {\tt mpdgangjobs}
help to manage parallel jobs.  There are a few others, and it is easy to write
new console commands as needed.
Finally,
in order that this infrastructure be available to support MPI programs or other
parallel tools, there needs to be a {\em client library\/} that
each application process may use to interact with its manager.

We do not specify how the daemons are started or connected, since the system
provides a number of alternatives, and the process need not be particularly
fast.  A console command is started by the user, either interactively or under
the control of a batch scheduler.  The daemons {\tt fork} and {\tt exec} the
managers, which use information given them by the daemons to connect
themselves into a ring, then {\tt fork} and {\tt exec} the clients.  The
startup messages traverse the ring quickly, so most {\tt fork}ing, {\tt
  exec}ing, and connecting takes place in parallel, leading to fast startup
even for large jobs.
\begin{figure}[htbp]
    \centerline{
      \epsfxsize=5.0in
      \epsfbox{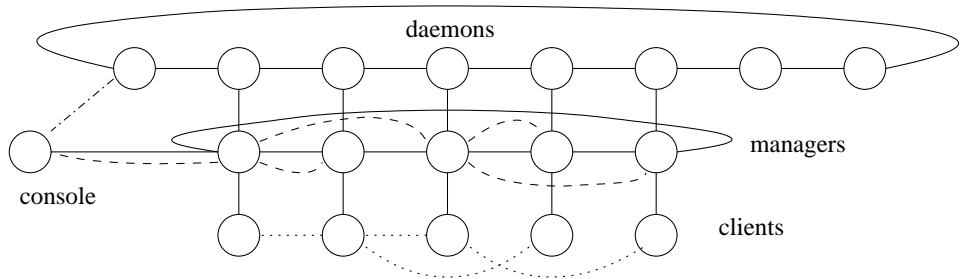}
    }
    \caption{Daemons with console process, managers, and clients}
    \label{fig:mpds-all}
\end{figure}
The situation is then as shown in Figure~\ref{fig:mpds-all}, where the clients
may be application MPI processes.  Solid lines represent sockets, except for
the vertical ones, which represent pipes.  The dashed lines represent the trees
of connections for forwarding {\tt stdout} and {\tt stderr}, and the dotted
lines represent {\em potential\/} connections among the client processes.  The
dot-dashed line is the original connection from console to local daemon on a
Unix socket, which is replaced during startup by the network connection to the
first manager.

\subsection{Interesting Features}
\label{sec:features}

Space restrictions prevent a complete description of all the features and
capabilities of the MPD system, but in this section we mention a few
highlights.

\begin{description}
\item[Security]
Whenever a process advertises a ``listener'' socket and accepts connections on
it, the possibility exists that an unknown or even malicious process will
connect.  This is particularly dangerous if the process accepting the
connection can start processes as the MPD daemon can.  We currently use
the ``challenge-response'' system described in~\cite{tanenbaum:networks}.
In the long run, we expect to modify this component of the system to use more
elaborate schemes and extend them to other connections such as client/gateway
authentication.  This will have little impact on the job startup speed, since
the daemon component startup is separate from job startup.

\item[Fault Tolerance]
If a daemon dies, this fact is detected and the ring is reknit.  This provides
a minimal sort of fault tolerance, since the ring remains intact.  A new MPD
daemon can be inserted in the ring where the old one was, but
this process is not (yet) automatic.

\item[Signals]
  Signals can be delivered to client processes by their managers.  We
  currently use this capability in three specific ways.  First, signals
  delivered to a console process are propagated to the clients, so that a
  parallel application as a whole can be suspended with {\tt cntl-Z},
  continued, and killed with {\tt cntl-C}, just as if it were a single
  process.  Second, in the {\tt ch\_p4mpd} device in the MPICH implementation
  of MPI, client processes can interrupt one another with requests to
  dynamically establish client-to-client connections.  Such requests go up
  into the manager ring from the originating client, around the ring to the
  manager of the target process, which signals its client.  Third, a separate
  console process can be used to implement a simple but effective gang
  scheduler.  Gang scheduling is the process of ensuring that all the
  processes of a parallel job are swapped in and scheduled to run at the same
  time across a collection of machines; it is particularly important for jobs
  that contain a large number of collective operations.  MPD can provide
  gang scheduling by using signals to pause and resume parallel jobs running on
  the same collection of machines.  This mechanism is also useful for pausing
  long-running parallel jobs to run short parallel utility programs.

\item[Support for MPI Implementations]
  MPD provdes a version of the BNR library described in Section~\ref{sec:bnr}.
  Version 1.2.1 of MPICH makes calls to this library to find out a process's
  rank, where other processes are and how to contact them, and so forth.
  
  \hspace{.16in} On clusters of SMPs, it is easy to specify that multiple
  processes are to be started on the same machine and share memory.
  Specifically, {\tt mpirun -np 180 -g 2 cpi} starts processes in groups of
  two and places in their environment a key that can be used to acquire
  group-attached shared memory and other information needed to set up
  multimethod communication for an MPI implementation.  We use this technique
  on the Chiba City cluster at Argonne, which has dual CPU nodes.

\item[Handling Standard I/O]
  Managers capture the {\tt stdout} and {\tt stderr} of their clients and
  forward them up a pair of binary trees of socket connections, each manager
  merging {\tt stdout} and {\tt stderr} from its client with that from each of
  its two children. A command line option tells the managers to provide a rank
  label on each line of output from their clients.
  A pedestrian but useful application of this feature is that it helps with
  programs that may not have been written to be parallel in the first place.
  Both standard output and error output are automatically identified with
  their source process without touching the original code.  The feature is also
  useful when invoking system utilities in ``task farm'' mode.  The command
\begin{verbatim}
    mpdmpexec -np 128 ps auxww | grep mpdman
\end{verbatim}
  finds (in conjunction with the use of {\tt hostname}) where {\tt mpd}'s have
  started managers.
  
\hspace{.16in} Standard input (to {\tt mpirun}, for example) by default is
delivered to the client managed by manager 0.  This seems to be what most MPI
users expect and what most MPI implementations do.  (The MPI standard does
not specify.)  However, control messages can be used to change this behavior
to direct {\tt stdin} to any specific client or broadcast it to all clients.

\item[Environment Variables]
  By default the {\tt DISPLAY} environment variable of the shell in which {\tt
    mpirun} is invoked is forwarded to the managers and set for the clients.
  This allows clients to use X graphics.  We plan to replace this non-scalable
  approach with one similar to the one used for {\tt stdout} and {\tt stderr}.
  Other environment variables can be specified on the command line for
  propagating to the application processes.

\item[Client Wrapping]
The semantics of the Unix {\tt fork} and {\tt exec} system calls provide 
useful benefits.  When a manager {\tt fork}s a client process, for example, it
first sets up the manager-client pipes for control messages and standard I/O.
The ``lower'' ends of these pipes are inherited by any process that the client
forks.  Thus, even though the client is not using any of the client library,
managers can manage clients that themselves run the ``real'' application
process.  We call this scheme {\em client wrapping}.  Thus
{\tt mpirun -np 16 nice -5 myprog}
lowers the priority of a parallel job to be run on one's colleagues'
workstations, and 
{\tt mpirun -np 16 pty myprog}
can be used when {\tt myprog} needs to be attached to a terminal (otherwise
our capture of {\tt stdin} and {\tt stdout} modifies their buffering behavior).
The program {\tt pty} is distributed with the MPD system.

\item[Putting It All Together]
  The combination of I/O management, especially redirection of {\tt stdin},
  line labels on {\tt stdout}, and client wrapping can be surprisingly
  powerful.  We have used these features of the MPD system to add an option to
  {\tt mpirun} that invokes {\tt gdb} as a client wrapper and dynamically
  redirects {\tt stdin}.  While {\tt mpirun -np 3 cpi} runs {\tt cpi} directly
  as an MPI job, {\tt mpirun -np 3 -d cpi} runs each {\tt cpi} process under
  the control of (wrapped by) the {\tt gdb} debugger. (Other sequential
  debuggers could be used, but are not yet supported.)  Thus multiple
  instances of {\tt gdb} are being run.  Output of the {\tt gdb}'s is labeled
  by process rank.  The ``{\tt (gdb)}'' prompts are intercepted by the {\tt
    mpirun} process and counted, so that it can issue an ``{\tt (mpigdb)}''
  prompt when one has been received from each process.  In addition, {\tt
    mpirun -d} uses the ``{\tt z}'' command (one of the few single letters not
  already claimed by {\tt gdb}) to redirect {\tt stdin} to a specific {\tt
    gdb} instance or to all processes.  Thus processes can be stepped and
  breakpoints can be set either collectively or individually, and collectively
  printing a variable will provide all values with rank labels. 
The following is an example of how this works:
\begin{small}
\begin{verbatim}
  donner% mpirun -np 3 -d cpi                  # default is stdin bcast
  (mpigdb) b 33                                # set breakpoint for all
  0: Breakpoint 1 at 0x8049eac: file cpi.c, line 33.
  1: Breakpoint 1 at 0x8049eac: file cpi.c, line 33.
  2: Breakpoint 1 at 0x8049eac: file cpi.c, line 33.
  (mpigdb) r                                   # run all
  2: Breakpoint 1, main (argc=1, argv=0xbffffab4) at cpi.c:33
  1: Breakpoint 1, main (argc=1, argv=0xbffffac4) at cpi.c:33
  0: Breakpoint 1, main (argc=1, argv=0xbffffad4) at cpi.c:33
  (mpigdb) n                                   # single step all
  2: 43           MPI_Bcast(&n, 1, MPI_INT, 0, MPI_COMM_WORLD);
  0: 39               if (n==0) n=100; else n=0;
  1: 43           MPI_Bcast(&n, 1, MPI_INT, 0, MPI_COMM_WORLD);
  (mpigdb) z 0                                 # limit stdin to rank 0
  (mpigdb) n                                   # single step rank 0
  0: 41               startwtime = MPI_Wtime();
  (mpigdb) n                                   # until caught up
  0: 43           MPI_Bcast(&n, 1, MPI_INT, 0, MPI_COMM_WORLD);
  (mpigdb) z                                   # go back to bcast
  (mpigdb) n                                   # single step all
                 ....                          # several times
  (mpigdb) n                                   # until interesting spot
  0: 52                   x = h * ((double)i - 0.5);
  1: 52                   x = h * ((double)i - 0.5);
  2: 52                   x = h * ((double)i - 0.5);
  (mpigdb) p x                                 # bcast print command
  0: $2 = 0.0050000000000000001                # 0's value of x
  2: $2 = 0.025000000000000001                 # 2's value of x 
  1: $2 = 0.014999999999999999                 # 1's value of x 
  (mpigdb) c                                   # continue all  
  0: pi is approximately 3.1416009869231249, Error  0.0000083333333318
  0: Program exited normally.
  1: Program exited normally.
  2: Program exited normally.
  (mpigdb) q                                   # quit     
  donner% 
\end{verbatim}
\end{small}

\item[Running the Daemons as Root]
  By default, the MPD daemons are run in ordinary user mode.  This is useful
  for development, but in production we do not wish the machines to fill up
  with {\tt mpd} processes being run by various users; we prefer to have only
  one {\tt mpd} per machine.  To this end the daemons can be configured to be
  run as root.  In this situation the console is a {\tt setuid} program that
  runs as root only to connect briefly to the local {\tt mpd}, then reassumes
  the user's user id, group id, and group memberships and sends these to the
  {\tt mpd}, so that the managers and clients run as the user in every way.
  In this mode the daemon is running as a ``true'' daemon, detached from any
  specific terminal and logging information and error messages to {\tt
  syslog}. 

\end{description}

\section{A General Process Manager Interface}
\label{sec:bnr}

One reason for creating MPD was that no existing process manager really had
the needed support for parallel jobs that we could use for MPICH.  One
research goal of the project was to determine a minimal set of services that
would need to be added to an existing commercial or open source process manager
in order to provide what was needed by a parallel library to implement MPI,
especially MPI-2, with its dynamic process creation and one-sided operations.
In this section we describe such an interface and how it is used by
MPICH and implemented by MPD.  The interface decouples MPD from MPICH,
allowing MPICH to be used with any process manager that implements this simple
interface and allowing other sorts of parallel systems besides MPICH to be
supported by the MPD runtime system.

In Section~\ref{sec:intro} we mentioned that we consider the scheduler,
process manager, and parallel programming library to be separate components of
a parallel environment.  We have tried to isolate and simplify the interface
between the process manager and the parallel programming library into a simple
specification called
BNR,\footnote{BNR stands for Bill, Brian, Nick, Rusty, and Ralph, who discussed
  it until it became as minimal as it is now.}
which will be reported on in more detail elsewhere.

BNR is the interface by which the parallel programming library (an example is
MPICH) obtains information from the process manager (an example is MPD) that
only the process manager initially knows, such as the rank in the parallel job
of an individual process and the information necessary for a process to
dynamically forge a connection with a process with a different rank.  It is
also the interface by which the parallel program library requests actions on
the part of the process manager.  A typical request would be to start
processes, either as part of MPI-2's {\tt MPI\_Comm\_spawn} or with {\tt
  mpirun} or {\tt mpiexec}.

The central component of BNR is a primitive database interface consisting of
job-local {\tt put}, {\tt get}, and {\tt fence} calls, by which processes can
place keyword=value data into the database, retrieve it by keyword, and
coordinate with the other processes in a parallel job to ensure that expected
data is deposited before it is accessed.  A few other commands also
are used to form dynamic connections.

We use this particular interface for scalability.  The use of the {\tt fence}
primitive permits a single synchronization operation after which all data that
has been {\tt put} before the {\tt fence} can be retrieved by a {\tt get}.
The {\tt get}, {\tt put}, and {\tt fence} calls are local to {\em groups\/} of
processes within jobs, which in an MPI implementation can correspond to MPI
groups.  They play an important role in libraries that use dynamic process
creation calls (such as {\tt MPI\_COMM\_Spawn}).

MPD implements the BNR interface by keeping the database distributed among the
managers for a job and routing data access requests around the manager ring as
necessary.  In the {\tt ch\_p4mpd} device in the current version of MPICH, we
use only this interface in the implementation, so that any process manager
implementing this interface can be used to manage MPICH programs without
knowing any MPICH internals.

\section{Experiments}
\label{sec:experiments}

Most development of MPD has been on workstation networks where startup of
32-process jobs on five workstations is virtually instantaneous, compared with
the approximately 1.5 seconds per process required by the {\tt ch\_p4} version
of MPICH.  An early test of the feasibility of using the ring topology showed
that a message could make 1024 hops around the ring in less than 0.4 seconds,
which gave us confidence that the ring would not impose scalability limits, at
least in the near term.  Recently we began experiments on Chiba City, a Linux
testbed for parallel computer science research~\cite{chibacity}.  We performed
one set of tests on 211 nodes connected by Fast Ethernet.  Because we were
interested only in process startup time, we tested execution of trivial
parallel jobs, for example,
\begin{verbatim}
    time mpirun -np 211 hostname
    time mpirun -np 422 -g 2 hostname
\end{verbatim}
We found that starting 211 processes (one on each node) and collecting the
{\tt stdout} output of {\tt hostname} took about 2 seconds to execute.
Starting twice as many processes (one for each CPU) took about 3.5 seconds,
including setting up the relatively complex {\tt stdout} tree
and collecting the output.  Sending a message around the ring of 211 MPD
daemons took only 0.13 seconds. 

MPD is now running on our Chiba City cluster in root mode, serving as an
experimental production process manager.  We have added facilities that allow
the MPD daemons to start jobs linked with Myricom's MPI implementation,
MPICH-GM, so that MPI jobs can be started with MPD and run over Chiba City's
Myrinet network.








\section{Future Development}
\label{sec:plans}

The existing MPD system, consisting of daemons, managers, console commands, and
client library, meets our goals of simplicity, robustness, and
scalability.  It is used for fast startup of MPI jobs and others on
systems with hundreds of machines.  The flexibility of its {\tt stdio} control
mechanism has provided unexpected benefits, such a ``poor man's'' parallel
debugger.  It meets our goals for the collective identity of a parallel job.

In the near term, we expect to use the system to implement the dynamic process
creation part of MPI-2 in MPICH.      
We also are working on a precise definition of how MPD will
interoperate with a full-featured scheduling system such as the Maui
scheduler~\cite{maui-scheduler}.  We believe that the MPD daemons can also
begin to provide more services, such as run-time performance monitoring.

In the long run, as machines grow from hundreds to thousands of nodes, our
rings of daemons and managers may have to grow into a more sophisticated
structure, such as rings of rings, in order to continue to provide fast
startup.  We anticipate that this can be done without substantially changing
the MPD design presented here.
We will also need a more sophisticated output merger in order to
provide scalable {\tt stdout}, for example for large-scale parallel debugging.

In summary, we are finding the MPD system already a useful contribution to
one's parallel programming environment and expect its applicability to
expand in the near future.  We also view its design as a valuable starting
point for future research into large-scale parallel job execution environments.

\bibliography{/home/MPI/allbib}

\begin{thebibliography}{10}

\bibitem{chibacity}
{Chiba City} home page.
\newblock http://www.mcs.anl.gov/chiba.

\bibitem{maui-scheduler}
{The Maui scheduler} home page.
\newblock http://maui-scheduler.mhpcc.edu/new\_doc, http://www.mhpcc.edu/maui.

\bibitem{baker96:clusters}
M.~A. Baker, G.~C. Fox, and H.~W. Yau.
\newblock Review of cluster management software.
\newblock {\em NHSE Review}, 1(1), May 1996.

\bibitem{Barak:1993:MDO}
Amnon Barak, Shai Guday, and Richard~G. Wheeler.
\newblock {\em The {MOSIX} distributed operating system: {Load balancing for
  UNIX}}, volume 672 of {\em Lecture Notes in Computer Science}.
\newblock Springer-Verlag, New York, 1993.

\bibitem{Beck99:harness}
Micah Beck, Jack~J. Dongarra, Graham~E. Fagg, G.~Al Geist, Paul Gray, James
  Kohl, Mauro Migliardi, Keith Moore, Terry Moore, Philip Papadopoulous,
  Stephen~L. Scott, and Vaidy Sunderam.
\newblock {HARNESS}: A next generation distributed virtual machine.
\newblock {\em International Journal on Future Generation Computer Systems},
  15(5/6), 1999.

\bibitem{lam}
Greg Burns, Raja Daoud, and James Vaigl.
\newblock {LAM}: An open cluster environment for {MPI}.
\newblock In John~W. Ross, editor, {\em Proceedings of Supercomputing Symposium
  '94}, pages 379--386. University of Toronto, 1994.

\bibitem{bgl00:mpd:pvmmpi00}
R.~Butler, W.~Gropp, and E.~Lusk.
\newblock A scalable process-management environment for parallel programs.
\newblock In Jack Dongarra, Peter Kacsuk, and Norbert Podhorszki, editors, {\em
  Recent Advances in Parallel Virutal Machine and Message Passing Interface},
  number 1908 in Springer Lecture Notes in Computer Science, pages 168--175,
  September 2000.

\bibitem{p4-paper}
Ralph Butler and Ewing Lusk.
\newblock Monitors, messages, and clusters: {T}he p4 parallel programming
  system.
\newblock {\em Parallel Computing}, 20:547--564, April 1994.

\bibitem{dqs-home-page}
{DQS} home page.
\newblock http://www.scri.fsu.edu/\char`\~pasko/dqs.html.

\bibitem{feitelson95:survey}
Dror~G. Feitelson.
\newblock {A Survey of Scheduling in Multiprogrammed Parallel Systems}.
\newblock Research report rc 19790 (87657), IBM T.J. Watson Research Center,
  Yorktown Heights, NY, February 1995.

\bibitem{foster99grid}
I.~Foster and C.~Kesselman, editors.
\newblock {\em The Grid: Blueprint for a New Computing Infrastructure}. Morgan
  Kaufmann, 1999.

\bibitem{pvmbook}
Al~Geist, Adam Beguelin, Jack Dongarra, Weicheng Jiang, Bob Manchek, and Vaidy
  Sunderam.
\newblock {\em {PVM}: {P}arallel Virtual Machine---A User's Guide and Tutorial
  for Network Parallel Computing}.
\newblock MIT Press, Cambridge, Mass., 1994.

\bibitem{Ghormley:1998:GGL}
Douglas~P. Ghormley, David Petrou, Steven~H. Rodrigues, Amin~M. Vahdat, and
  Thomas~E. Anderson.
\newblock {GLUnix}: {A} {Global Layer Unix} for a network of workstations.
\newblock {\em Soft\-ware---Prac\-tice and Experience}, 28(9):929--961, July
  1998.

\bibitem{Gropp:1998:MPI2Book}
William Gropp, Steven Huss-Lederman, Andrew Lumsdaine, Ewing Lusk, Bill
  Nitzberg, William Saphir, and Marc Snir.
\newblock {\em {MPI}---The Complete Reference: Volume 2, The {MPI}-2
  Extensions}.
\newblock MIT Press, Cambridge, MA, 1998.

\bibitem{SUT}
William Gropp and Ewing Lusk.
\newblock Scalable {U}nix tools on parallel processors.
\newblock In {\em Proceedings of the Scalable High-Performance Computing
  Conference}, pages 56--62. {IEEE Computer Society Press}, 1994.

\bibitem{gropp-lusk-doss-skjellum:mpich}
William Gropp, Ewing Lusk, Nathan Doss, and Anthony Skjellum.
\newblock A high-performance, portable implementation of the {MPI}
  {M}essage-{P}assing {I}nterface standard.
\newblock {\em Parallel Computing}, 22(6):789--828, 1996.

\bibitem{loadleveler}
IBM.
\newblock {\em Loadleveler: Using and Administering}, version 2 release 1
  edition, November 1998.
\newblock SA22-7311-00.

\bibitem{Litzkow88}
M.~J. Litzkow, M.~Livny, and M.~W. Mutka.
\newblock Condor -- {A} hunter of idle workstations.
\newblock In {\em Proc. 8th Intl. Conf. on Distributed Computing Systems},
  pages 104--111, San Jose, Calif., June 1988.

\bibitem{Migliardi:1999:PEH}
M.~Migliardi and V.~Sunderam.
\newblock {PVM} emulation in the {Harness} metacomputing system: {A} plug-in
  based approach.
\newblock In J.~J. Dongarra, E.~Luque, and Tomas Margalef, editors, {\em Recent
  advances in parallel virtual machine and message passing interface: 6th
  European {PVM}\slash {MPI} Users' Group Meeting, Barcelona, Spain, September
  26--29, 1999: Proceedings}, volume 1697 of {\em Lecture Notes in Computer
  Science}, pages 117--124, Berlin, 1999. Spring{\-}er-Ver{\-}lag.

\bibitem{pbs-home-page}
{PBS} home page.
\newblock http://pbs.mrj.com/.

\bibitem{platform-home-page}
{Load Sharing Facility (LSF)}.
\newblock http://www.platform.com.

\bibitem{Pruyne:1996:ICP}
J.~Pruyne and M.~Livny.
\newblock Interfacing {Condor} and {PVM} to harness the cycles of workstation
  clusters.
\newblock {\em Future Generation Computer Systems}, 12(1):67--85, May 1996.

\bibitem{Snir:1998:MPI2Book}
Marc Snir, Steve~W. Otto, Steven Huss-Lederman, David~W. Walker, and Jack
  Dongarra.
\newblock {\em {MPI}---The Complete Reference: Volume 1, The {MPI} Core, {\rm
  2nd edition}}.
\newblock MIT Press, Cambridge, MA, 1998.

\bibitem{tanenbaum:networks}
Andrew~S. Tanenbaum.
\newblock {\em Computer Networks}.
\newblock Prentice Hall, third edition, 1996.

\end{thebibliography}
\bibliographystyle{plain}

\end{document}